# Layer-dependent electromechanical response in twisted graphene moiré superlattices

Hanhao Zhang[1,#], Yuanhao Wei[2,#], Yuhao Li[1,3,#,*], Shengsheng Lin[1], Jiarui Wang[1], Takashi Taniguchi[4], Kenji Watanabe[5], Jiangyu Li[6], Yi Shi[1], Xinran Wang[1,7,8], Yan Shi[2,*], Zaiyao Fei[1,3,*]

1. National Laboratory of Solid-State Microstructures, School of Electronic Science and Engineering and Collaborative Innovation Center of Advanced Microstructures, Nanjing University, Nanjing 210093, Jiangsu, China
2. State Key Laboratory of Mechanics and Control of Mechanical Structures, Nanjing University of Aeronautics and Astronautics, Nanjing 210016, Jiangsu, China
3. National Key Laboratory of Spintronics, Nanjing University, Suzhou 215163, China
4. Research Center for Materials Nanoarchitectonics, National Institute for Materials Science, 1-1 Namiki, Tsukuba 305-0044, Japan
5. Research Center for Electronic and Optical Materials, National Institute for Materials Science, 1-1 Namiki, Tsukuba 305-0044, Japan
6. Guangdong Provisional Key Laboratory of Functional Oxide Materials and Devices, Department of Materials Science and Engineering, Southern University of Science and Technology, Shenzhen 518055, Guangdong, China
7. School of Integrated Circuits, Nanjing University, Suzhou 215163, Jiangsu, China
8. Suzhou Laboratory, Suzhou 215123, Jiangsu, China

[#]These authors contributed equally to this work.
[*]Correspondence to: zyfei@nju.edu.cn, yuhao@nju.edu.cn and yshi@nuaa.edu.cn

**ABSTRACT:** The coupling of mechanical deformation and electrical stimuli at the nanoscale has been a subject of intense investigation in the realm of materials science. Recently, twisted van der Waals (vdW) materials have emerged as a platform for exploring exotic quantum states. These states are intimately tied to the formation of moiré superlattices, which can be visualized directly exploiting the electromechanical response. However, the origin of the response, even in twisted bilayer graphene (tBLG), remains unsettled. Here, employing lateral piezoresponse force microscopy (LPFM), we investigate the electromechanical responses of marginally twisted graphene moiré superlattices with different layer thicknesses. We observe distinct LPFM amplitudes and spatial profiles in tBLG and twisted monolayer-bilayer graphene (tMBG), exhibiting effective in-plane piezoelectric coefficients of 0.05 pm/V and 0.35 pm/V, respectively. Force tuning experiments further underscore a marked divergence in their responses. The contrasting behaviors suggest different electromechanical couplings in tBLG and tMBG. In tBLG, the response near the domain walls is attributed to the flexoelectric effect, while in tMBG, the behaviors can be comprehended within the context of piezoelectric effect. Our results not only provide insights into electromechanical and corporative effects in twisted vdW materials with different stacking symmetries, but may also offer a way to engineer them at the nanoscale.

**INTRODUCTION**

Electromechanical coupling effects, such as piezoelectric, flexoelectric and their reverse effects, are highly desirable for applications in sensing, actuating and energy harvesting[1-4]. Generally, for a crystalline dielectric, the induced electrical polarization upon mechanical deformation can be phenomenologically formulated by the constitutive equation[5,6] $P_i = e_{ijk}\epsilon_{jk} + \mu_{ijkl}\frac{\partial \epsilon_{jk}}{\partial x_l}$. The first term describes the conventional piezoelectric effect and is exclusive to non-centrosymmetric materials. The second term corresponds to the flexoelectric effect, which can appear in any materials. At the nanoscale, the flexoelectric effect may become significant in the presence of a pronounced strain gradient[7]. An example of this is the interfacial electromechanical coupling that usually arises at the interfaces or boundaries between two different materials[8-10]. While the interfacial piezoelectricity and flexoelectricity are critical for micro/nano electromechanical systems, understanding of the interface effects is hindered by the lack of well-defined interfaces and efficient tuning knobs.

In parallel, twisting two graphene sheets has emerged as a way of manipulating their electronic properties, leading to the discovery of abundant correlated and topological electronic states[11-23]. For marginally twisted graphene, lattice reconstruction of the moiré superlattices would result in alternating commensurate stacking domains[24-29] (*e.g.* AB/BA domains for tBLG and ABA/ABC domains for tMBG). The stacking dislocations are confined to the domain walls (lateral interfaces), across which the stacking order changes abruptly. Thus, significant strain and strain gradient are anticipated near these domain walls[30], making twist graphene moiré superlattice a platform for exploring interfacial electromechanical couplings. However, even the fundamental piezoelectric coefficients in these systems remain undetermined. Furthermore, while LPFM can be utilized to visualize the moiré superlattices[31,32], the origin of the electromechanical coupling is yet to be fully understood. For instance, while tBLG is thought to exhibit LPFM signals near domain walls due to the flexoelectric effect, quantitative analyses of the spatial profiles appear inconsistent with the model[31]. In twisted multilayer graphene, the inversion symmetry of individual stacking domains may be broken, implying that the piezoelectric effect could also contribute to the electromechanical response near the domain walls.

In this study, we methodically investigate the electromechanical responses of marginally tBLG and tMBG moiré superlattices and find stark contrast between them. The effective in-plane piezoelectric coefficients for tBLG and tMBG are determined to be 0.05 pm/V and 0.35 pm/V, respectively. Moreover, we demonstrate the potential for engineering electromechanical responses in twisted multilayer graphene through applied force.

**RESULTS**

We focus on tBLG and tMBG with a tiny twist angle of below 0.1 degree. The experimental setup is illustrated in Figure 1a, in which the projected in-plane deformation is recorded and mapped as an electric field is applied (single frequency LPFM, unless otherwise specified). The sample is comprised of an exposed twisted graphene sitting atop a hexagonal boron nitride (h-BN) substrate. For some samples, they are also contacted by metal electrodes through few-layer graphene for conductive atomic force microscopy (cAFM) measurements (see Methods and SI-1 for details).

Figure 1b shows the schematic lattice structure of a hybrid tBLG-tMBG sample (htG-I, see Figure S2a for the optical image) in the rigid model, which is fabricated by tearing and stacking a graphene flake that consists of both mono- and bi-layer graphene. A large-area LPFM phase image of the sample is provided in Figure 1c (see Figure S2b for the amplitude image). On the top right and bottom left corners, we can see well-developed tBLG and tMBG moiré superlattices, in which clear LPFM contrasts are observed near the domain walls. The top left and bottom right insets show the enlarged phase images of tBLG and tMBG, respectively. Line cuts of the phases and amplitudes are provided in Figure 1d. The amplitude and spatial profiles are very different for tBLG and tMBG in the following two aspects. First, the LPFM amplitude difference of tBLG is about an order of magnitude smaller than that of tMBG. Second, in the vicinity of a domain wall, one can see two distinct extrema with different phases in tBLG, while tMBG seems to develop a broad peak. Similar observations are also made in another hybrid sample (htG-II) in Figure S3.

The moiré wavelengths of the tMBG above is ~200 nm, significantly smaller than that of tBLG (~600 nm). In Figure 2, we show fine scans of the same tBLG region and a nearby tMBG region (moiré wavelength ~400 nm). Figure 2a&d are the vectorially decomposed (decoupled from the background) phase and amplitude images[32] (see Figure S4 for the raw images). Line cuts of the phases and amplitudes across a domain wall is plotted in Figure 2b&e. Clearly, there are two separate LPFM extrema of opposite polarities near domain walls of tBLG, which is reminiscent of the in-plane electrical polarization generated by the flexoelectric effect[8,31,33,34]. The feature maintains when the tBLG sample is rotated by ~90°, as shown in Figure S5. The contrasting line cuts clearly demonstrate the different electromechanical responses of tBLG and tMBG. LPFM images of other tBLG and tMBG samples can be found in Figure S6&S7. We would like to point out that while the broad peak in tMBG persists for different drive frequency, the observation of opposite polarities in tBLG is very sensitive to the drive frequency of the cantilever (see SI-2, Figure S8&S9 for details), which is probably related to the rather weak electromechanical response in tBLG[31].

To obtain the in-plane piezoelectric coefficients, we adopt the dual AC resonance tracking (DART) LPFM, and measure the deformations at different drive voltages. Figure 2c&f show the averaged deformations at three selected domain walls as a function of the drive voltage for tBLG (sample tBLG-I) and tMBG (sample tMBG-I), respectively, along with their linear fits. The fitted slopes are different for different domain walls, depending on the relative angle between the domain wall and the cantilever axis[31,32]. When the domain wall is perpendicular to the cantilever axis, the slope would correspond to the effective in-plane piezoelectric coefficients. From the insets to Figure 2c&f, the coefficients are determined to be 0.05 pm/V for tBLG and 0.35 pm/V for tMBG (see SI-2, Figure S10&S11 for details). The extracted value for tMBG is about an order of magnitude smaller than the in-plane piezoelectric coefficient ($d_{11}$) of monolayer $MoS_2$[35].

Although line cuts across tMBG domain walls do not show LPFM signals of opposite polarities, a faint secondary peak / shoulder seems to develop (Figure 2e). To examine it, we apply different forces to the twisted graphene sample through the AFM tip while performing LPFM measurements. The raw phase images at representative forces on tBLG and tMBG are shown in Figure 3a&d (images at other forces can be found in Figure S12&S13), respectively. In contrast to the immunity to the applied force for tBLG, LPFM signals near the domain walls of tMBG split into two separate peaks at large forces. In most tMBG samples, the split lines are not perfectly parallel, a feature that appears to be associated with the curvature of the domains. Figure 3b&e plot the line cuts of the phases across the same domain walls as marked by the pink arrow line in Figure 3a&d, respectively.

The spatial profiles are fitted by bimodal Gaussian functions. While the peak and dip positions for the Fano-like curve of tBLG are almost independent of the applied force, the peak separation of tMBG increases with the applied force. The extracted peak-to-peak (or dip) separation (open cyan circles) and span (open pink triangles) for tBLG and tMBG as a function of the applied force are provided in Figure 3c&f, respectively. We also normalize the widths to the moiré wavelengths and fit them to lines. The slope of the linear fit for tMBG is more than 10 times larger than that of tBLG. The splitting of peaks for tMBG may imply that the domain wall regions broaden with the applied force. This is corroborated by the force-dependent cAFM measurements on similar tMBG moiré superlattices (SI-3 and Figure S14).

Except for tBLG and tMBG, we also studied electromechanical responses of twisted monolayer-trilayer graphene (tMTG) and twisted double bilayer graphene (tDBG) moiré superlattices (see Figure S15-S17 for details). Overall, the spatial profiles and force dependences resemble those of tMBG, and the amplitudes (under the same drive voltage and force) exhibit small variations. The drastic difference between tBLG and twisted multilayer graphene (tMBG, tMTG and tDBG) suggest that the electromechanical coupling of twisted multilayer graphene is dominated by another mechanism than the flexoelectric effect. We propose it is the piezoelectric effect.

**DISCUSSION**

Microscopically, charge redistribution in response to the shear strain (gradient) can give rise to electric dipoles near the domain walls of twisted graphene. Inversely, the application of an external electric field would modify the local electronic structures and electric dipoles, which in effect reshape the lattice reconstruction, resulting in detectable mechanical deformations. Phenomenologically, the strain $\epsilon_{ij}$ generated by applying electric field $E_k$ can be expressed as[31,36] $\epsilon_{ij} = d_{ijk}E_k$. Here the piezoelectric coefficient $d_{ijk}$ can be related to the electrical polarization ($P_v$) through[37] $d_{ijk} = C_{ijlm}^{-1}\eta_{lu}Q_{mkuv}P_v$, where $C_{ijlm}^{-1}$ is elastic coefficient, $\eta_{lu}$ is the dielectric constant, $Q_{mkuv}$ is the electrostriction coefficient. In fact, charge redistribution enhanced/induced piezoelectric responses on domain wall boundaries have been observed at 2H-1T' phase boundaries of few-layer MoTe$_2$[38] and biaxial-strained monolayer graphene[39] before.

From the constitutive equation, the piezoelectric contribution to the induced electric polarization is proportional to the strain field, while the flexoelectric contribution is proportional to the strain gradient field. Here, the strain field is $\epsilon_{jk} = \frac{1}{2}(\frac{\partial u_j}{\partial x_k} + \frac{\partial u_k}{\partial x_j})$, where $\vec{u}$ is the interlayer displacement field. For an ideal moiré domain wall along the y axis (Figure 4a), the in-plane displacement is along the domain wall direction ($\vec{u} = u_y \hat{y}$), i.e. a shear-type dislocation. Thus, only $\epsilon_{xy} = \epsilon_{yx} = \frac{1}{2}\frac{\partial u_y}{\partial x}$ demand evaluation. In Figure 4b, we sketch profiles of the displacement, shear strain and normal derivative of the shear strain across the domain wall for tBLG, where the profile of the displacement is adopted from Ref.[40,41]. Similar profiles are also expected for tMBG. Without going to details of the electromechanical coefficients, the flexoelectric term mimics the in-phase LPFM profile of tBLG (Figure 2b&S18), while the piezoelectric term appears like that of tMBG at small forces (Figure 2e&S18). The faint secondary peak in tMBG may arise from the slightly different displacement field near the saddle point or SP stacking region. Given the three-fold rotational symmetry and the alignment of deformations along domain wall boundaries, we sketch the 2D deformation images in Figure 4c&d, originating from the shear strain and normal derivative of the shear strain, respectively. Under a moderate force, domain walls of tBLG almost stay intact, resulting in the same strain and strain gradient field. However, for tMBG, the domain walls would broaden as the force increases, reshaping the displacement field. In this case, split LPFM peaks of the same polarity may appear as illustrated in Figure 4e. Note the phenomenological model does not take the strain inhomogeneity into account, which could be a future direction.

We could also appreciate the different behaviors of tBLG and tMBG from the symmetry perspective. Near the SP stacking of marginally tBLG, the local inversion symmetry is broken by the strain gradient. In-plane electric dipoles are allowed on either side of the domain walls, oriented in opposite directions because of the symmetry between AB and BA stackings. The overall pattern of electric dipole preserves the global six-fold rotational symmetry[42]. For tMBG, across the domain walls, the stacking order changes from ABA to ABC. During the transition, the local inversion and the three-fold rotational symmetries are necessarily broken irrespective of the strain gradient. In this scenario, the piezoelectric effect can dominate over the flexoelectric effect, and in-plane electric dipoles of the same orientations can form near the domain walls. Additionally, given the relatively large domain wall boundary width, we anticipate a restricted strain gradient compared to atomically sharp interfaces. These may explain the weaker electromechanical responses (LPFM amplitudes and the effective in-plane piezoelectric coefficients) observed in tBLG compared to tMBG.

In reality, the electromechanical responses in twisted multilayer graphene may also be complicated by other factors such as substrate effects[31], impurities and defects in graphene[43,44], the stacking energy imbalance and work function differences across different stacking domains. The response to external force in tMBG may point to the effect of the asymmetry between ABA and ABC stackings, which deserves further exploration.

**CONCLUSIONS**

In summary, twisted graphene moiré superlattices represent a pristine platform for investigating electromechanical couplings at the nanoscale with exceptional tunability. Our findings can facilitate similar investigations in other moiré superlattices. The intricate interplay of electromechanical coupling and the diverse electronic states of matter in moiré systems[45-49] may not only advance our understanding of the electronic phenomena but also promise for the development of innovative technologies with tailored electronic and mechanical functionalities.

**METHODS**

**Sample fabrication.** All devices were fabricated using a modified 'tear-and-stack' method, similar to Ref.[16]. Typical optical images of twisted graphene devices are shown in Figure S2.

**Atomic force microscopy measurements.** Three AFM modes were adopted in this work, *i.e.*, LPFM, DART LPFM and cAFM, all performed on an Oxford Instruments Asylum Research MFP-3D Origin AFM. ASYELEC-01-R2 probes coated with 5-nm Ti and 20-nm Ir were used in all modes with a spring constant of 2.8 N/m. Typical free resonance frequency is ~75 kHz and LPFM contact resonance frequency is ~780 kHz. The applied force between tip and sample is regulated by the voltage of setpoint. More details of AFM measurements can be found in SI-1.

**Data processing.** To determine the effective in-plane piezoelectric coefficient, under different applied forces, the decoupled LPFM amplitudes at selected domain walls are collected, and Gaussian fittings are performed on them. The mean values and standard deviations of the fittings correspond to the projected in-plane deformations and error bars in Figure 2c&f, respectively. In Figure 3c&f, the span is taken as the width at 5% of the maximum values of the left and right peaks (or dips), the error bars represent the 90% confidence intervals.

## ASSOCIATED CONTENT

**Supporting Information**

The Supporting Information is available free of charge *via* the internet at http://pubs.acs.org.

Single frequency LPFM and DART LPFM measurements, Frequency dependence of LPFM results, Force-dependent cAFM measurement are presented here.

**Notes**

The authors declare no competing interests.

## ACKNOWLEDGEMENTS

We thank Menghao Wu and Konstantin Shapovalov for useful discussions. This work is supported by the National Key Research and Development Program of China (2021YFA0715600), the National Natural Science Foundation of China (12274222), the National Science Foundation of Jiangsu Province (BK20220756) and the Postgraduate Research & Practice Innovation Program of Jiangsu Province (KYCX24_0145 and KYCX24_0147). Yan.S. is supported by the National Natural Science Foundation of China (12072150) and the National Natural Science Foundation of China for Creative Research Groups (51921003). K.W. and T.T. acknowledge support from the JSPS KAKENHI (21H05233 and 23H02052) and the World Premier International Research Center Initiative (WPI), MEXT, Japan.

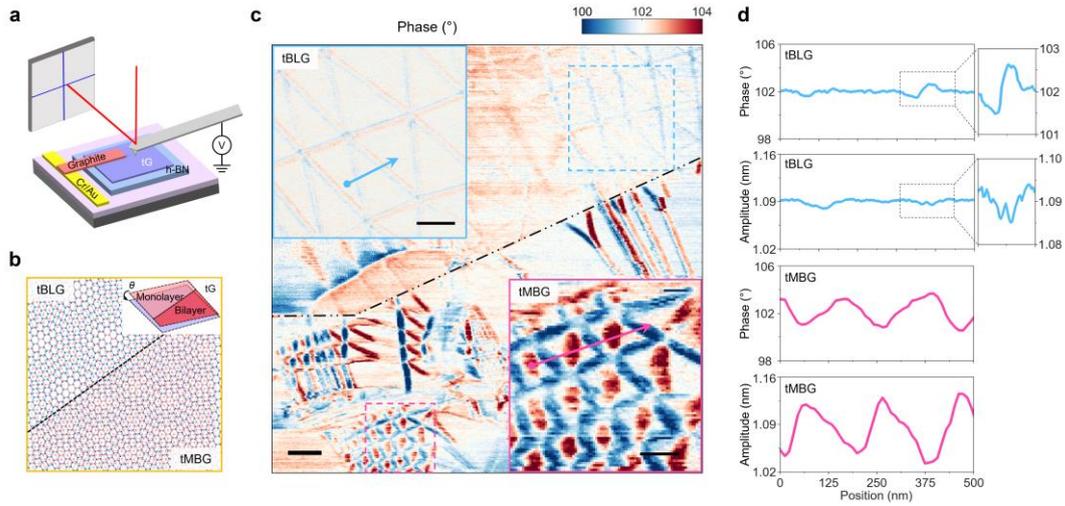

**Figure 1. LPFM mapping in a hybrid tBLG-tMBG sample (htG-I).** (**a**) Schematic of the LPFM setup and sample structure. (**b**) Schematic of the hybrid sample. The upper left and bottom right are tBLG and tMBG regions, respectively. (**c**) Phase images of htG-I. The insets are enlarged phase images of tBLG (blue dashed square) and tMBG (red dashed square). Scale bars: 400 nm for the main figure and the upper left inset, 200 nm for the bottom right inset. (**d**) Line cuts of the phases and amplitudes for tBLG and tMBG along blue and red arrow lines in the inset of (c), respectively. The insets zoom in a domain wall region of tBLG.

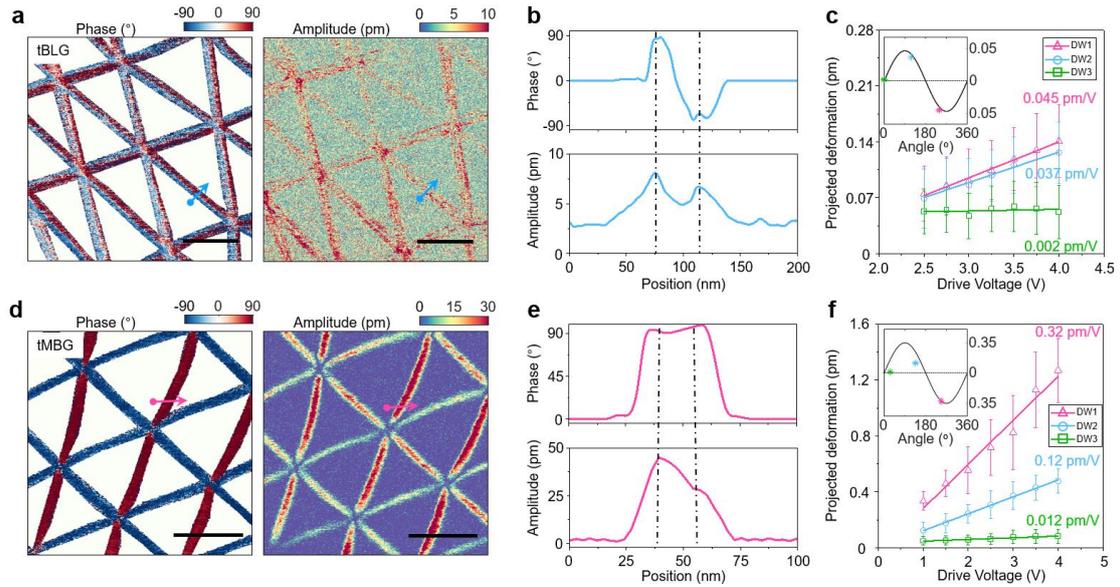

**Figure 2. Decoupled LPFM of tBLG and tMBG with similar moiré wavelengths.** (**a**) Decoupled phase and amplitude images of tBLG. (**b**) Line cuts of the decoupled phase and amplitude along the blue arrow lines in (a). (**c**) Averaged LPFM amplitude (projected deformation) at three selected domain walls (red triangle, blue circle and green squares) of tBLG as a function of the AC drive voltage, along with the linear fits. The error bars are standard deviations of the amplitude. The slopes of the linear fits, which are measured in-plane piezoelectric coefficients for three domain walls in one scan, are 0.045 pm/V, 0.037 pm/V and 0.002 pm/V, respectively. Inset: sinusoidal fit of the three slopes versus the cantilever axis angle relative to the domain wall. The maximum value of the function corresponds to the effective in-plane piezoelectric coefficient, which is 0.05 pm/V for tBLG. (**d**) Decoupled phase and amplitude maps of tMBG. (**e**) Line cuts of the decoupled phase and amplitude in (d) along the pink arrow lines. (**f**) Drive voltage dependence of DART LPFM amplitudes for three selected tMBG domain walls. Similar to the data processing for tBLG in (c), the effective in-plane piezoelectric coefficient is determined to be 0.35 pm/V. Scale bars in (a) and (d) are 400 nm.

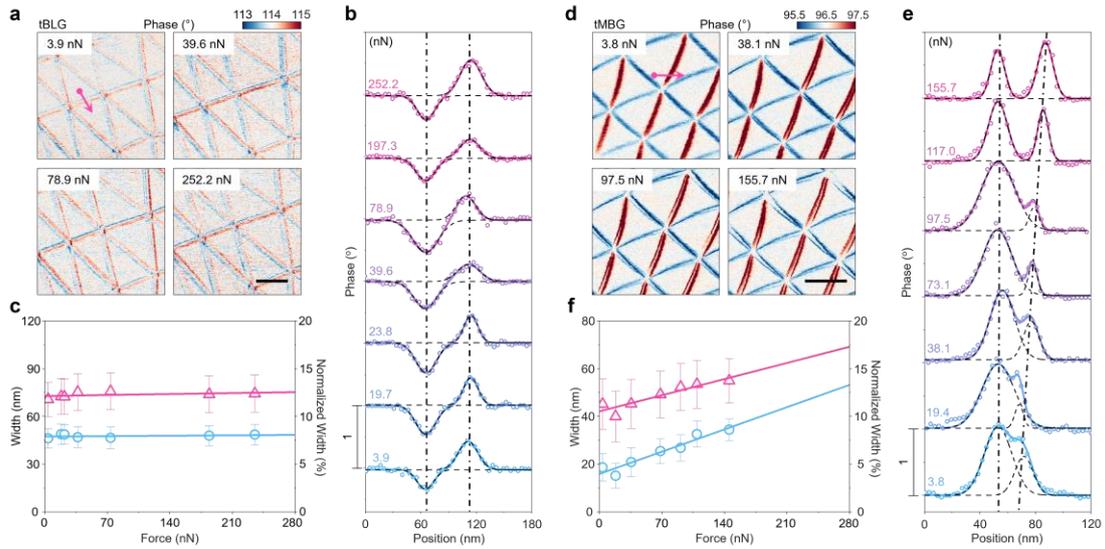

**Figure 3. Force tuning of the electromechanical responses near the domain walls.** (**a**) LPFM phase images of tBLG under selected applied force. (**b**) Line cuts of the phase for tBLG along the pink arrow line in (a) under different applied forces from 3.9 nN to 252.2 nN. The open circles represent the measured data, while the dashed and solid lines correspond to the individual and combined Gaussian fits. Sequential curves are offset by 1° for clarity. (**c**) Peak-to-peak separation (cyan open circles) and span (pink open triangles) of tBLG as a function of the applied force, along with the linear fits. The slopes are 0.0035 nm/nN (cyan) and 0.0071 nm/nN (pink), respectively. (**d**) LPFM phase images of tMBG under selected applied forces. (**e**) Line cuts of the phase for tMBG along the pink line in (d) under different applied forces from 3.8 nN to 155.7 nN. (**f**) Peak-to-peak separation (cyan open circles) and span (pink open triangles) of tMBG as a function of the applied force, along with the linear fits. The slopes are 0.124 nm/nN (cyan) and 0.902 nm/nN (blue), respectively. Scale bars for (a) and (d) are 400 nm.

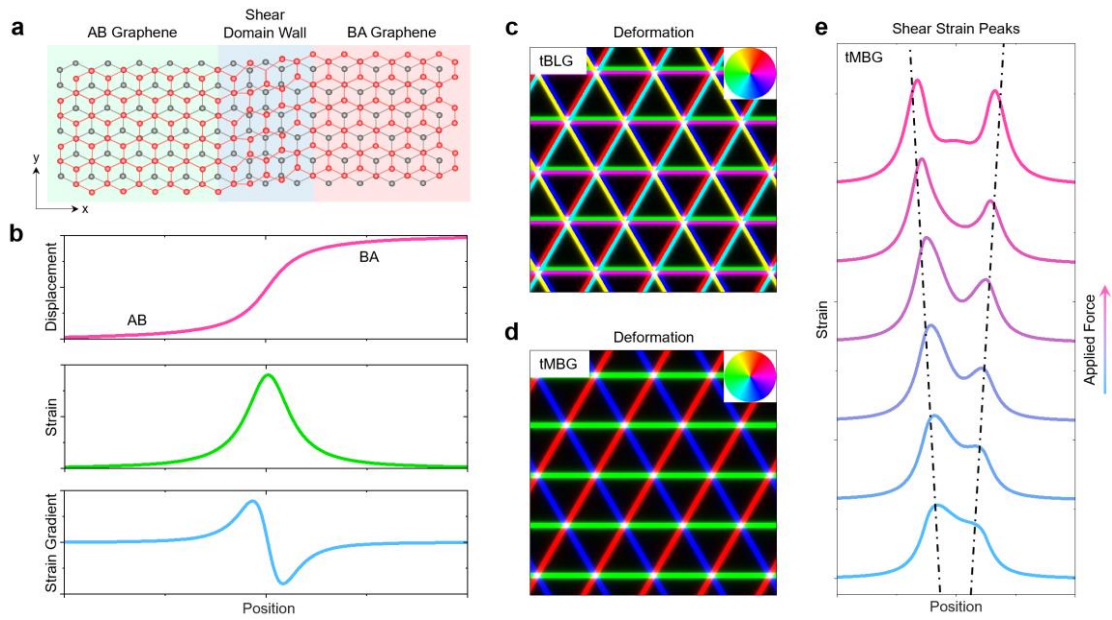

**Figure 4. Analysis of flexoelectric and piezoelectric effects in graphene moiré superlattices.** (**a**) Schematic of a tBLG moiré domain wall. The red and black hexagonal grids represent the upper and lower graphene layers, respectively. (**b**) Spatial profiles of the displacement ($u_y$), shear strain ($\epsilon_{xy}$) and normal derivative of the shear strain ($\partial \epsilon_{xy}/\partial x$) fields across the shear domain wall in a. (**c** and **d**) Schematic deformation images that are proportional to the shear strain and shear strain gradient, respectively. The color denotes the direction of deformation as in the insets. (**e**) Splitting of shear strain peaks near a domain wall in tMBG as a function of the applied force.